\documentclass[runningheads]{llncs}

\usepackage[utf8]{inputenc}
\usepackage[T1]{fontenc}
\usepackage{float}
\usepackage{graphicx}
\usepackage{listings}
\usepackage{xcolor}
\usepackage{listings-ext}
\usepackage{amsmath}
\usepackage{amssymb}

\usepackage[hidelinks]{hyperref}
\usepackage[noabbrev,capitalize]{cleveref}

\usepackage{algpseudocodex}

\usepackage{booktabs}

\usepackage{todonotes}

\definecolor{codegreen}{rgb}{0,0.6,0}
\definecolor{codegray}{rgb}{0.5,0.5,0.5}
\definecolor{codepurple}{rgb}{0.58,0,0.82}
\definecolor{backcolour}{rgb}{0.95,0.95,0.92}

\lstdefinestyle{mystyle}{
  backgroundcolor=\color{backcolour}, commentstyle=\color{codegreen},
  keywordstyle=\color{magenta},
  numberstyle=\tiny\color{codegray},
  stringstyle=\color{codepurple},
  basicstyle=\ttfamily\footnotesize,
  breakatwhitespace=false,         
  breaklines=true,                 
  captionpos=b,                    
  keepspaces=true,                 
  numbers=left,                    
  numbersep=5pt,                  
  showspaces=false,                
  showstringspaces=false,
  showtabs=false,                  
  tabsize=2,
  escapeinside={(*}{*)},
  extendedchars=false
}

\begin{document}

\title{Quantifying Liveness and Safety of Avalanche's Snowball}
%
\author{Quentin Kniep\inst{1} \and Maxime Laval\inst{2} \and Jakub Sliwinski\inst{1} \and Roger Wattenhofer\inst{1}}
\authorrunning{Q. Kniep et al.}

\institute{ETH Zurich\\Zurich, Switzerland \\
    \email{\{qkniep,jsliwinski,wattenhofer\}@ethz.ch} \and
EPFL\\Lausanne, Switzerland \\
    \email{maxime.laval@epfl.ch}}


\maketitle

\begin{abstract}
    This work examines the resilience properties of the Snowball and Avalanche protocols that underlie the popular Avalanche blockchain.
    We experimentally quantify the resilience of Snowball using a simulation implemented in Rust, where the adversary strategically rebalances the network to delay termination.
    
    We show that in a network of $n$ nodes of equal stake, the adversary is able to break liveness when controlling $\Omega(\sqrt{n})$ nodes.
    Specifically, for $n=2000$, a simple adversary controlling $5.2\%$ of stake can successfully attack liveness. When the adversary is given additional information about the state of the network (without any communication or other advantages), the stake needed for a successful attack is as little as $2.8\%$.
    
    We show that the adversary can break safety in time exponentially dependent on their stake, and inversely linearly related to the size of the network, e.g. in 265 rounds in expectation when the adversary controls $25\%$ of a network of 3000. 
    
    We conclude that Snowball and Avalanche are akin to Byzantine reliable broadcast protocols as opposed to consensus.
\end{abstract}
%

\section{Introduction}

The Avalanche protocol~\cite{consensus_whitepaper} advertises exceptional performance in terms of transaction throughput and latency.
The Avalanche blockchain based on the protocol has certainly gained significant attention and support within the cryptocurrency community, as evidenced by the remarkable market capitalization of its native token amounting to \$10B\footnote{\url{https://coinmarketcap.com} (Accessed: June 19 2024)}. The media prominence and monetary value firmly place Avalanche among the most popular and successful blockchain systems.

The protocol is built on a simple mechanism that operates by repeatedly sampling random nodes of the network in order to gauge the system's support of a given decision and confirm transactions.
Conceptually, the underlying Snowball protocol can be compared to a voting process for a binary choice concerning a transaction.
The protocol description promises to swiftly converge to a final decision from a network state initially divided equally between two alternatives.
With the aid of a directed acyclic graph (DAG), Avalanche forms a partial order of transactions instead of the total order that is established by usual blockchain protocols, like Bitcoin~\cite{bitcoin} or Ethereum~\cite{ethereum}.
Thus when transactions on Avalanche are accepted, validators are able to execute them in different orders based on their current view of the DAG, as long as those transactions are not causally dependent on each other.
In theory, such structure can allow for a higher degree of parallelism in the transaction confirmation process, which can lead to a higher throughput than traditional blockchain protocols.

The ideas that form Avalanche stand in contrast to other Proof-of-Stake and BFT-based consensus protocols such as Ethereum 2.0.
While the whitepaper~\cite{consensus_whitepaper} claims excellent resilience, it only proves the protocol's liveness in presence of up to $\mathcal{O}(\sqrt{n})$ malicious parties, where $n$ represents the total number of validators (or stake supply).
However, usually Proof-of-Stake protocols ensure the upper bound resilience of $\frac{n}{3}$ in partial synchrony.

Another detail that stands out in the description of Avalanche, is how it defines its guarantees with respect to ``virtuous'' transactions, i.e. assuming there's no conflicting alternative in the system. Remarkably, broadcast-based payment systems \cite{consensusnumber,fastpay} are inherently reliant on such an assumption, and as such are fundamentally weaker than consensus protocols.

The lack of clarity about the Avalanche family of protocols begs the question: how resilient Snowball and Avalanche really are? Does the unusual consideration of ``virtuous'' transactions indicate a fundamental limitation?


\subsubsection{Our Contribution}
We examine the resilience properties of the Snowball and Avalanche protocols.

We experimentally exhibit the resilience of Snowball against attacks from adversarial nodes.
Our simulation showcases that in a system of $n$ nodes (or stake supply), the adversary can indefinitely halt the Snowball protocol when controlling a stake of $\Omega(\sqrt{n})$, or less than 2\% in some experimental scenarios.
Furthermore, we examine a strategy for an adversary to violate safety by getting a single validator to finalize an output distinct from the rest of the network.
The expected duration of the safety attack depends exponentially on the stake controlled by the adversary, and is inversely linear to the size of the network.
For example, at $25\%$ adversarial stake in a network of 3000, safety can be violated after 265 rounds in expectation.

We discuss how these considerations translate to the Avalanche protocol based on Snowball.

Finally, we draw parallels between Avalanche and broadcast-based payment systems, and conclude that Avalanche is fundamentally weaker than usual consensus protocols.

\section{Background}

Avalanche's blockchain platform consists of three distinct built-in blockchains:
The Exchange Chain (X-Chain), the Contract Chain (C-Chain) and the Platform Chain (P-Chain)~\cite{avalanche_platform}.
\\ \\
The \textbf{X-Chain} is responsible for processing simple transactions on the network, such as transfers of the native AVAX token.
It is based on the Avalanche protocol with the DAG that runs multiple instances of the Snowball algorithm and only partially orders transactions. 
\\ \\
The \textbf{C-Chain} is responsible for executing general smart contracts compatible with the Ethereum Virtual Machine (EVM). In contrast to the X-Chain, C-Chain uses the Snowman protocol which ensures a total order of all transactions.
\\ \\
The \textbf{P-Chain} processes various platform-level operations, such as creation of new blockchains and sub-networks, validator (de-)registration, or staking operations. It also uses Snowman.
\\ \\
The Avalanche protocol introduced in the Ava Labs whitepaper, which is also mainly marketed and presented in online materials, is used as the basis of X-Chain. Interestingly, the Snowman protocol, which supports the C-Chain and P-Chain, is almost absent from documentation and marketing, and remains outside the scope of this work.

\subsection{Validators}

Participants in the Avalanche protocol are called validators or nodes. Validators following the protocol are called honest. As a blockchain protocol, Avalanche aims to be resilient to validators deviating from the protocol, which are called malicious, or collectively as the adversary.

Avalanche employs a Proof-of-Stake mechanism to control the ability of malicious validators joining the system. Validators need to acquire AVAX tokens (2,000 minimum) and deposit them using the Avalanche platform to actively participate in the agreement process. Validators are associated with, and weighted by, the amounts of deposited tokens, called their \emph{stake}. Typically, Proof-of-Stake blockchains aim to be resilient to the adversary that is able to acquire a stake smaller than 1/3 of the total tokens (which is the theoretical maximum in harsh network conditions).

%



\subsection{UTXO Model}
Avalanche uses the Unspent Transaction Output (UTXO) model, as initially introduced in Bitcoin~\cite{bitcoin}.
In the model, a transaction contains a set of inputs, a set of outputs, and a digital signature.
Each input of a transaction corresponds to a specific output from a previous transaction.
Transactions are issued by users, processed by the system, and as a result are accepted or rejected by the system.

Two transactions including the same input are \emph{conflicting}, and only one transaction from such a pair can be accepted by the system.

The balance of a user is determined by the set of outputs transferred to that user in previously accepted transactions and not yet used as inputs for newer transactions.
A valid transaction is also signed with keys corresponding to the relevant inputs.

In contrast to most blockchains such as Bitcoin, Avalanche does not necessitate a total order of all transactions.
Instead, transactions in Avalanche form a directed acyclic graph (DAG) resulting in a partial order.
A transaction $\emph{tx}{'}$ depends on $\emph{tx}$ if $\emph{tx}^\prime$ consumes an output of \emph{tx}.
In this case, every validator needs to process \emph{tx} before processing $\emph{tx}^\prime$.
Validators can execute transactions that are not dependent on each other in any order.


\section{Snowball}
\label{sec:snowball}
The Snowball protocol serves as the foundational component of the Avalanche blockchain.
It is based on continuously querying random sets of $k$ validators regarding their current ``approval'' regarding a transaction, denoted as $T$.

When performing a query on $k = 20$ nodes within a Snowball instance, the selection probability of a node is proportional to the stake of the node.
Intuitively, the influence of validators in validating transactions, quantified by the probability of them being queried, is determined by their stake.

Validators maintain a confidence value for each binary choice:
Blue if they prefer to accept transaction $T$, Red if they reject transaction $T$.
When a validator queried $k$ other nodes and saw at least $\alpha$ for either Red or Blue, we say that this color received a \emph{chit}, and the confidence value for that color is incremented by one.
When queried, a validator will either respond Blue if the confidence value for Blue is higher, or Red if the confidence value for Red is higher.
A color is accepted by a node if for at least $\beta$ consecutive rounds of querying it received a chit.
The logic of Snowball is illustrated in \Cref{fig:snowball}.

\begin{figure}[!t]
    \begin{algorithmic}
        \State $k, \alpha, \beta \gets 20, 15, 20$ \Comment{Protocol parameters}
        \Statex
        \Function{Snowball}{$V, v_\mathsf{self}, c_\mathsf{init}$}
            \State $c_\mathsf{pref} \gets c_\mathsf{init}$
            \State $c_\mathsf{last} \gets c_\mathsf{init}$
            \State $confidence \gets [0, 0]$
            \State $counter \gets 0$
            \Statex
            \While{$counter < \beta$}
                \State $V_\mathsf{query} \gets \textsc{SampleVal}(V \setminus \{v_\mathsf{self}\}, k)$ \Comment{Weighted by stake with replacement.}
                \State $R \gets \textsc{Query}(V_\mathsf{query})$ \Comment{Query each with $v_\mathsf{self}$, $c_\mathsf{pref}$; $R$ is multiset of responses.}
                \For{$i \in \{0, 1\}$}
                    \If{$|[r \in R\ |\ r=i]| \ge \alpha$} 
                        \If{$c_\mathsf{last} \neq i$}
                            \State $counter \gets 0$
                        \EndIf
                        \State $confidence[i] \gets confidence[i] + 1$
                        \If{$confidence[i] > confidence[1-i]$}
                            \State $c_\mathsf{pref} \gets i$
                        \EndIf
                        \State $c_\mathsf{last} \gets i$
                        \State $counter \gets counter + 1$
                    \EndIf
                \EndFor
            \EndWhile
            \State \Return $c_\mathsf{last}$
        \EndFunction
        \Statex
        \Function{RespondToQuery}{$querier, c_\mathsf{querier}$}
            \If{$c_\mathsf{pref} = \bot$}
                \State $c_\mathsf{pref} \gets c_\mathsf{querier}$
            \EndIf
            \State \Return $c_\mathsf{pref}$
        \EndFunction
    \end{algorithmic}
    \caption{Snowball algorithm.}
    \label{fig:snowball}
\end{figure}

\subsection{Safety}
Intuitively, safety properties can be understood as ``bad'' things not happening. In our context, the main safety property is ensuring that two honest nodes cannot perceive two conflicting transactions as accepted. The Avalanche whitepaper outlines the definition of safety as follows:  \\ \\ \textbf{P1. Safety:} When decisions are made by any two honest nodes, they decide on conflicting transactions with negligible probability ($\le \varepsilon$).

Here $\varepsilon$ represents the safety failure probability, with the specific value dependent on the maximum number $f$ of adversarial nodes, which is not explicitly stated in the formal definition provided by the Avalanche whitepaper.

\subsection{Liveness}
Liveness refers to the continued operation of the system. In our context, liveness mainly refers to ensuring that all honest nodes eventually decide to accept or reject a transaction within a reasonable time frame.
 \\ \\
According to the whitepaper, Avalanche has the following liveness guarantees: 
\\ \\
\textbf{P2. Liveness (Upper Bound):} Snow protocols terminate with a strictly positive probability within $t_{max}$ rounds.  \\ \\
\textbf{P3. Liveness (Strong Form):} If $f \in \mathcal{O}(\sqrt{n})$, then the snow protocol terminates with high probability $(\geq 1 - \varepsilon$) in $\mathcal{O}(\log(n))$ rounds.
\\ \\ 
However, it is specified later in the whitepaper that P2 holds only under the assumption that initially, one proposal has at least $\frac{\alpha}{k}$ support in the network, for which there is no guarantee.

\section{Simulation}
\label{chapter:simulation_results}

To test resilience of Snowball, we perform a local simulation of the protocol using a Rust implementation \cite{ourcode}. As the base implementation of Snowball (c.f. \autoref{fig:snowball}) we use the \texttt{avalanche-consensus} Rust library\footnote{https://crates.io/crates/avalanche-consensus},
which is a translation of the Snowball Go code that is part of the official Avalanche implementation\footnote{https://github.com/ava-labs/avalanchego} and is maintained by the Ava Labs team.

The simulation involves a network of multiple honest nodes executing the protocol correctly, aiming to achieve agreement on a binary decision. Malicious nodes collude to perform the considered attacks. In our experimental scenarios, the stake is equally divided among validators.

\subsection{Network Assumptions}

Distributed protocols might require various network reliability assumptions to work correctly. Many blockchain protocols guarantee safety in harsh network conditions, such as those of partially synchronous models, where messages can be greatly delayed.

In our simulation we make the strongest network reliability assumptions possible, where every message arrives with the same, known latency. We deny the adversary any communication advantage whatsoever, including advantages often practically achievable by an attacker in the real world, such as performing queries faster, or performing more queries.

In synchronous rounds, nodes query other nodes, as described by the Snowball protocol. Between rounds, the nodes update their preferred color with which they respond to the queries.

\subsection{Adversary Information}

To perform the attacks, the adversary needs information about the other nodes' preferred color. We call the adversary \emph{naive} if the adversary simply queries honest nodes in line with the protocol and updates his estimation of the colors preferred by the honest nodes according to the query results.

We also consider an adversary that possesses accurate information about the numbers of nodes preferring Red/Blue in the current round, and call that adversary \emph{informed}. 

\subsection{Liveness Attack}
\label{sec:liveness_attack}

When attacking the liveness property, the adversary aims to delay the decision of honest nodes by keeping the network split equally between Red and Blue.
The attack strategy we consider is straightforward.
When the adversary is queried, it responds with the color that is less preferred among all honest validators.
By doing so, the adversary aims to bring the network split between the Red and Blue decisions closer to the even 50-50 split.
This is shown in \autoref{fig:liveness-attack}.

\begin{figure}[!t]
    \begin{algorithmic}
        \State $n, f$ \Comment{Network parameters.}
        \State $\mu_\mathsf{estimate}$ \Comment{Current estimate of network-wide preference towards 1.}
        \Statex
        \Function{RespondToQuery}{$v_\mathsf{query}, c_\mathsf{querier}$}
            \If{$\mu_\mathsf{estimate} < 0.5$}
                \State \Return 1
            \EndIf
            \State \Return 0
        \EndFunction
    \end{algorithmic}
    \caption{Adversary strategy for liveness attack.}
    \label{fig:liveness-attack}
\end{figure}




For a given experimental scenario, we consider the attack successful if in more than 5 out of 10 simulation runs, no validator has terminated with a decision after 100,000 rounds. We note that if a round of querying took about 1 second, 100,000 rounds would correspond to over a day.

\begin{figure}[h!]
    \includegraphics[width=\columnwidth]{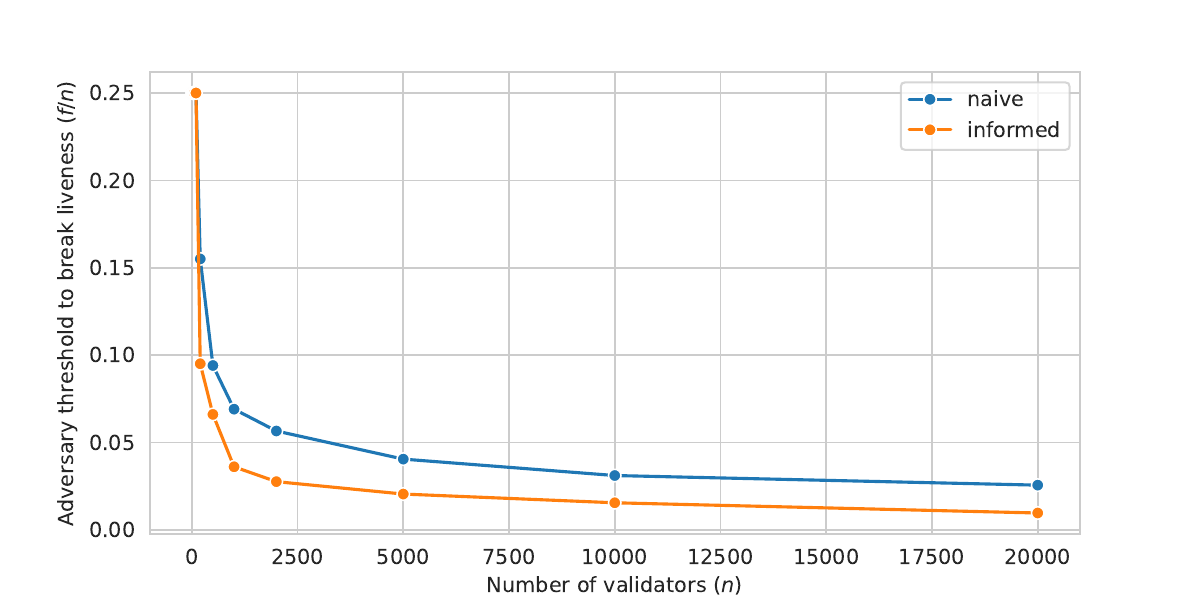}
    \caption{Minimum fraction of stake needed by the adversary to successfully attack liveness, plotted against the number of nodes in the network with equal stake.
    }
    \label{fig:min_adv_liveness}
\end{figure}

We perform binary search with respect to the adversary stake to find the minimal fraction of total stake for which the adversary is successful.
\autoref{fig:min_adv_liveness} shows the minimum percentage of stake the adversary needs to attack the liveness of the protocol.
It can be seen to decrease significantly with increasing number of total nodes in the network, showing the sub-linear security bound.

\subsection{Safety Attack}\label{section:safety_attack}
In this scenario, the adversary aims to break the safety property of the protocol by causing some honest nodes to accept conflicting transactions.

The adversarial strategy we consider starts with maintaining a modified liveness attack.
While the honest nodes are divided between Red and Blue, denote by $\mu$ the fraction of honest nodes that prefer Red. Then, the fraction of honest nodes that prefer Blue is $1-\mu$.
The adversary attempts to keep the numbers of honest nodes preferring Red and Blue close to the $\mu:Red$, $1-\mu:Blue$ split by replying to queries with colors that sway the honest nodes towards this split.
Additionally, the adversary chooses a set of honest nodes to queries of which the adversary responds exclusively with Red.
By employing this approach, the attacker can significantly increase the likelihood of the targeted nodes finalizing with the color Red after some time,
while at the same time keeping the rest of the network from deciding in either direction.
Once some targeted node accepts Red, the adversary replies to all queries with the color Blue, such that the rest of the network accepts Blue.
As a result, the adversary produces a safety violation, as the targeted node decides differently to the rest of the network.

\autoref{fig:safety-attack} describes the adversarial strategy for the considered safety attack.

\begin{figure}[!t]
    \begin{algorithmic}
        \State $n, f$ \Comment{Network parameters.}
        \State $V_\mathsf{target}$ \Comment{Validators targeted in the safety attack.}
        \State $\mu_\mathsf{estimate}$ \Comment{Current estimate of network-wide preference towards 1.}
        \State $\mu_\mathsf{target}$ \Comment{Target split to maintain before finalization.}
        \State $fin$ \Comment{Indicator if some targeted validator has finalized.}
        \Statex
        \Function{RespondToQuery}{$v_\mathsf{query}, c_\mathsf{querier}$}
            \If{$fin$}
                \State \Return 0
            \ElsIf{$v_\mathsf{query} \in V_\mathsf{target}$}
                \State \Return 1
            \EndIf
            \Statex
            \If{$\mu_\mathsf{estimate} < \mu_\mathsf{target}$} \Comment{Otherwise, continue with regular liveness attack.}
                \State \Return 1
            \EndIf
            \State \Return 0
        \EndFunction
    \end{algorithmic}
    \caption{Adversary strategy for safety attack.}
    \label{fig:safety-attack}
\end{figure}

%
%
%

\subsection{Safety Attack Analysis}
Consider the attack where a single node is targeted. Denote the number of adversarial nodes by $f$ and the number of nodes in total by $n$.
Assuming that the adversary can maintain the honest nodes split of $\mu:Red$, $1-\mu:Blue$, in expectation we observe the following:
when the targeted node queries, it receives a percentage of $\mu (1-\frac{f}{n}) + \frac{f}{n}$ responses for Red, while other nodes receive $\ge \mu (1-\frac{f}{n})$ fraction of responses for Red, depending on the adversary. 
For example, with 30\% adversary stake and a split of 69.4\% Red and 30.6\% Blue among honest nodes,
the targeted node has a probability $p = 0.694 \cdot 0.7 + 0.3 = 0.7858$ of receiving a Red response when querying.
Consequently, the targeted node can finalize Red with some probability, which eventually occurs.

Once this happens, the adversary replies to all queries with Blue. Since $\mu (1-\frac{f}{n}) = 0.694 \cdot 0.7 = 0.4858 < 0.5$, it is very likely that all honest nodes flip to Blue and later accept Blue.
In general, if $\mu (1-\frac{f}{n}) < 0.5$, the adversary still has the ability to sway the network towards accepting Blue.

We now compute the probability that the targeted node converges to Red, given that it sees an average proportion of $\mu (1-\frac{f}{n}) + \frac{f}{n}$ of responses in favor of Red when querying.
Recall that when querying $k = 20$ other nodes, a validator increments its successive success counter, denoted as ``counter'', only if a color receives at least $\alpha = 15$ votes, and if this color is the same as the currently preferred color. Otherwise, the success counter is reset to 0.

Let the random variable $X$ denote the number of participants who prefer Red in a sample of size $k = 20$. We want to calculate the probability distribution $P(X \ge \alpha) = 1 - P(X < 15)$.
We can model this using a binomial distribution with parameters $p = \mu (1-\frac{f}{n}) + \frac{f}{n}$ for the targeted node and $p = \mu (1-\frac{f}{n})$ for the other nodes.
Thus, we have:
\[ P(X \ge \alpha) = 1 - P (X< \alpha) = 1 - F(\alpha - 1,k,p) = 1 -\sum_{i=0}^{\alpha-1} \binom{k}{i} p^{i} (1-p)^{k-i} \]

Here, $F(\alpha - 1, k, p)$ represents the cumulative distribution function (CDF) of the binomial distribution. 
For our previous example, where $\mu = 0.694$, for honest nodes other than the attack target, we can calculate the probability $P(X \ge \alpha)$, which represents the chance of reaching the $\alpha$ majority threshold for the color Red when querying.
Plugging in the probability to receive a response supporting Red in a single round $p = 0.4858$ from above,
we get $P(X \ge \alpha) = 1 - F(14, 20, 0.4858) \approx 0.015$.  
On the other hand, the targeted node that has a probability $p = 0.7858$ to get a Red response, and so $p_{\alpha} = P( X \ge \alpha) = 1 - F(14, 20, 0.7858) \approx 0.756$.
This means that our targeted node has a $ p_{\alpha} \approx 75.6\%$ chance of reaching the $\alpha$ majority for Red when querying $k = 20$ other nodes,
whereas other honest nodes only have a $1.5\%$ chance of the same.
Consider the expected number of iterations needed to obtain $\beta = 20$ consecutive successes of reaching the $\alpha = 15$ majority for a color.
Let $X_{\beta}$ represent the number of trials required to achieve $\beta$ consecutive successes, with the probability of one success being $p_{\alpha}$.
From \cite{drekic2021number}, we can use the following formulas:
\[ \mathbb{E}[X_{\beta}] = \frac{1 - p_{\alpha} ^ \beta}{(1 - p_{\alpha}) p_{\alpha} ^ \beta  } \]
\[ \mathbb{V}ar[X_{\beta}] = \frac{1 - (2\beta + 1)(1 - p_{\alpha})p_{\alpha} ^{\beta} - p_{\alpha} ^{2 \beta + 1}}{(1 - p_{\alpha}) ^ 2 p_{\alpha} ^ {2 \beta}  } \]
For the example where $p_{\alpha} = 0.756$ for the targeted node, we obtain:
\[ \mathbb{E}[X_{20}] \approx 1095 \]
\[ \sigma = \sqrt {\mathbb{V}ar[X_{20}] } \approx 1078 \]

On average, the targeted node needs to query 1,095 times with a standard deviation of 1,078. We confirmed the expected results experimentally.

\begin{table}
    \centering
    \setlength{\tabcolsep}{6pt}
    \begin{tabular}{llllll}
    \toprule
    Adversary & Honest Split & $p_{\alpha}$ & $ \mathbb{E}[X_{20}]$ & $\sigma$ & $\approx\mathbb{E}[X_{20}^{1000}]$ \\ 
    \midrule
    30\% & 69.4\% - 30.6\% & 0.756 & 1,095   & 1,078   & 20 \\
    25\% & 64.8\% - 35.2\% & 0.560 & 245,562 & 245,544 & 265 \\
    20\% & 60.7\% - 39.3\% & 0.364 & 9.4e+8  & 9.4e+8  & 940,000 \\
    10\% & 54.0\% - 46.0\% & 0.101 & 8.4e+19 & 8.4e+19 & 8.4e+16 \\
    5\%  & 51.2\% - 48.8\% & 0.043 & 2.5e+27 & 2.5e+27 & 2.5e+24 \\
    \bottomrule
    \end{tabular}
    \vspace{4pt}
    \caption{Summary of the expected safety attack results for different percentages of adversarial stake and corresponding stable network splits. The second column shows the maximally imbalanced but stable split of honest validators that the adversary is able to maintain. $\mathbb{E}[X_{20}^{1000}]$ is the expected length of the safety attack when 1000 nodes are targeted.}
    \label{tbl:safety-attack}
\end{table}

The effectiveness of the attack can be greatly increased by targeting a large number of nodes rather than just one. Let $X_{\beta}^k$ be the number of trials required for any target node among $k$ targeted nodes to achieve $\beta$ consecutive successes. Assuming the $\beta$ successes to be equally probable to conclude in every round after 19, and assuming the constant network split maintained by the adversary, the expected number $\mathbb{E}[X_{\beta}^k]$ is $\frac{\mathbb{E}[X_{\beta} - 19]}{k}+19$. We have simulated some experimental scenarios, such as targeting 1000 nodes with $n = 3000$ and the adversarial stake of $f = 750$, where the results matched our expectation.

With increasing total number of nodes $n$, the adversary can target more nodes. While in our experiments we successfully targeted over $0.3n$ of $n = 3000$ nodes, future work is needed to understand how big the share of targeted nodes can be in an optimal strategy and with increasing $n$. In summary, the strength of the attack corresponds exponentially to the share of the adversary stake. On the other hand, the expected required duration of the attack is inversely linear to the overall number of nodes $n$, as the number of targeted nodes can increase roughly linearly with $n$. \autoref{tbl:safety-attack} summarizes the effectiveness of the safety attack for adversaries of different strengths.



\section{Avalanche Protocol}
In this section, we explain how the Avalanche protocol builds on Snowball to incorporate optimizations and additional features.

\subsection{DAG}
\label{sec:dag}


To enhance the throughput and enable parallel processing of transactions,
the Avalanche protocol builds a directed acyclic graph (DAG) for transactions, instead of a linear chain.
Each transaction is represented as a node in the DAG.
Furthermore, transactions in the DAG are interconnected through parent-child relationships:
A transaction $T$ refers to older transactions known as its parents $\mathsf{Parents}(T)$.
We denote the parent relation $T^{\prime} \in \mathsf{Parents}(T)$ by $T^{\prime} \leftarrow T$.
If $T''$ is reachable by parent links from $T$, we say that $T''$ is an ancestor of $T$, or $T'' \in \mathsf{Ancestors}(T)$, and that $T$ is a descendant of $T''$, or $T \in \mathsf{Descendants}(T'')$.


\subsection{Vertex}
In order to limit the coordination overhead, a node in the Avalanche DAG is not an individual transaction but rather a batch of transactions known as a \emph{vertex}.
A vote for a vertex is considered a vote for all transactions contained within that vertex.
This allows Avalanche to facilitate efficient queries, while still maintaining confidence levels and a conflict set for each individual transaction.

When a vertex is accepted, all transactions within it are accepted.
When a vertex is rejected, valid transactions in that vertex may be batched into a new vertex, by removing the non-preferred transactions that resulted in the vertex getting rejected.
When a node creates a vertex $V$, it chooses parents for $V$ that are currently preferred.

When a user submits a payload transaction $tx$, a node creates a transaction $T \langle tx,\mathcal{D} \rangle$ for that payload.
It includes the payload $tx$, along with the set of UTXO IDs that will be consumed if the transaction is accepted, and the list $\mathcal{D}$ of dependencies on which this transaction relies.
Each dependency must be accepted before this transaction can be accepted.
The node then batches this transaction $T \langle tx,\mathcal{D} \rangle$ with other pending transactions into a vertex.
The node assigns one or more parents to this vertex, allowing it to be added to the DAG.
We define an Avalanche transaction $T$ as preferred if it is the preferred transaction in its conflict set $P_T$.
In other words, if transaction $T$ has the highest confidence among other conflicting transactions.
Each node $u$ calculates the confidence value for each transaction $T$ denoted by $d_{u}[T]$.
This confidence value is defined as the sum of the chits received by $T$ and all its descendants \cite{consensus_whitepaper}:
$ d[T] =\sum_{ T^\prime \in \mathcal{T}_{u} : T \in \mathsf{Descendants}(T^\prime)}  c_{u,T^{\prime}} $.
Here, $\mathcal{T}_u$ represents all the transactions currently known by node $u$ in its view of the DAG, and $c_{u,T^{\prime}}$ represents the chit received by transaction $T^{\prime}$.
$c_{u,T}$ can only take two values: 0 or 1.
Node $u$ queries transaction $T$ only once, as the votes on the descendants of $T$ also serve as queries and votes on $T$.
Specifically:
\[c_{u,T}=\begin{cases}
1 & \text{transaction $T$ received a chit when $u$ queried for it}\\
0 & \text{otherwise} 
\end{cases}
\]
As a reminder, receiving a chit for transaction $T$ means that node $u$ received an approval rate of at least $\alpha = 15$ when it queried $k = 20$ other nodes to determine if $T$ was their preferred transaction.
The confidence value of $T$ (and thus its status as accepted or rejected) is then updated based on the queries made on its descendants.

We say that a transaction $T$ is strongly preferred if $T$ is preferred and all its ancestors are also preferred in their respective conflict sets. 
An Avalanche transaction $T$ is considered virtuous if it conflicts with no other transactions or if it is strongly preferred.
Consequently, a virtuous vertex is a vertex where all its transactions are virtuous.
Similarly, a preferred or strongly preferred vertex is one where all its transactions are preferred or strongly preferred, respectively.
The parents of a vertex are randomly chosen from the \emph{virtuous frontier} set $\mathcal{VF}$, which consists of the vertices at the frontier of the DAG that are considered virtuous:
\[ \mathcal{VF} = \{ \ T \in \mathcal{T} \ \mid \mathsf{virtuous}(T) \ \land \neg \ \mathsf{virtuous}(T^{\prime}) \ \forall T^{\prime} \in \mathcal{T} : T \leftarrow T^{\prime} \} \]
The notation $\mathsf{virtuous}(T)$ indicates that $T$ is virtuous.
In other words, $\mathcal{VF}$ is the set of vertices that are virtuous, and have no virtuous children.

\subsection{From Snowball to Avalanche}
The Avalanche protocol runs a Snowball instance on the conflict set of each transaction $T$ once a node hears about a new transaction that gets appended to the DAG.
This means that when a new transaction $T$ is received, a validator will query k other random nodes to determine if $T$ is their preferred transaction.
The queried nodes will respond positively only if transaction $T$ and its ancestors in the DAG are also their preferred transactions within their respective conflict sets.
Instead of querying a Snowball instance for each individual transaction, Avalanche batches transactions into a vertex and instantiates a Snowball instance for that vertex, checking if all the transactions within that vertex and its ancestors are valid.

When a node is queried about the preference of transaction $T$ and its ancestors, it provides not just a binary vote as in Snowball, but rather responds with its entire virtuous frontier $\mathcal{VF}$ based on its local view.
This allows the respondents to specify which ancestors are not preferred if $T$ is not strongly preferred.
The querying node $u$ collects the virtuous frontier of the $k$ queried nodes.
For each virtuous frontier $\mathcal{VF}^\prime$ sent by a node $w$ as a vote, we add the transactions $T^\prime$ from $\mathcal{VF}^\prime$ and the ancestors of $T^\prime$ to a set $\mathcal{G}[T,w]$, which represents the positively reported transactions of $w$ when asked to vote for $T$.
We then count how many times node $w$, when queried for $T$, has acknowledged a transaction $T^{\prime}$ as virtuous, and store this in the counter $ack[T,T^{\prime}]$.
We then run a Snowball instance for every $ack[T,T^{\prime}]$:
If $ack[T,T^{\prime}$] received more than $\alpha$ votes it indicates that the $\alpha$ majority of the $k$ queried validator agree that $T^{\prime}$ is preferred.
We then increase the consecutive counter for $T^{\prime}$ if it was also the preferred transaction in the last vote.
The above procedure of voting on a vertex containing a single transaction can be generalized for vertices containing multiple transactions.

Finally, there are two ways in which a vertex $V$, and consequently all the transactions it contains $T \in V$, can be accepted, provided that all the ancestors of $V$ have also been accepted.
The first way is if none of its transactions $T \in V$ conflict with any other transactions, and the vertex $V$ received $\beta_{1}$ consecutive successes.
In this case, the vertex and all its transactions are accepted by node $u$.
The second way is if some transactions $T \in V$ have other transactions in their conflict sets, and the vertex $V$ receives $\beta_{2}$ consecutive successes.
In this case, the node accepts the vertex $V$ and all its transactions.
The Avalanche protocol denotes $\beta_{1}$ as \texttt{betaVirtuous} and $\beta_{2}$ as \texttt{betaRogue}, and naturally $\beta_{1}$ < $\beta_{2}$.

\subsection{Liveness Attack}
Suppose that two transactions (both with accepted virtuous ancestors)
$T$ batched in vertex $V$ and $T^{\prime}$ batched in vertex $V^{\prime}$ are conflicting.
Recall that the Snowball liveness attack consisted of a strategy where the adversary tried to ensure that the split between Red and Blue was always close enough to 50\% each.
Here, the approach is similar, except that we have to ensure that, on average, 50\% of the network has $V$ in their virtuous frontier or as an ancestor of their virtuous frontier, and the other 50\% of the nodes have $V^{\prime}$ in their virtuous frontier or as an ancestor of their virtuous frontier.
The binary attack can be transposed to one where the adversaries responds with the virtuous frontier $\mathcal{VF}$, with $V$ an ancestor of the nodes in $\mathcal{VF}$, or responds with the virtuous frontier $\mathcal{VF}^{\prime}$, with $V^{\prime}$ an ancestor of the nodes in $\mathcal{VF}^{\prime}$.
The intuition behind this attack is that half of the nodes will adopt a virtuous frontier that contains vertex $V$ as a virtuous vertex, and the other half of the nodes will adopt a virtuous frontier that contains $V^{\prime}$ as a virtuous node.
At every iteration of the loop, the adversary needs to maintain those two conflicting virtuous frontiers $\mathcal{VF}$ and $\mathcal{VF}^{\prime}$, grow the DAG such that some new valid vertices are appended to the conflicting $\mathcal{VF}$ and $\mathcal{VF}^{\prime}$, and respond accordingly with either one of the virtuous forests using the same technique that was used for the Snowball liveness attack.

\subsection{Safety Attack}
The safety attack from Snowball to Avalanche can be transposed in the same way as was explained above for the liveness attack.
Similarly, we will try to maintain a network split that does not converge:
$\mu$ of nodes will prefer a virtuous frontier $\mathcal{VF}$ that contains v as an ancestor, and $\nu$ of nodes will prefer a virtuous frontier $\mathcal{VF}^{\prime}$ that contains $v^{\prime}$ as an ancestor.
For one targeted node, the adversary will respond exclusively with the virtuous frontier $\mathcal{VF}$ instead of trying to maintain a split.
This way, we can make the targeted node to accept vertex $v$ while all the other nodes are still undecided.
Once this is done, the adversary can unanimously respond with virtuous frontier $\mathcal{VF}^{\prime}$ to make the rest of the nodes accept $v^{\prime}$ in order to break safety.
Such an attack can be instantiated by any adversary that creates conflicting (double spending) transactions $T$ and $T^{\prime}$, batch them in nodes $v$ and $v^{\prime}$ and conducts the attack to make some nodes accept T, and some other nodes accept $T^{\prime}$ thus resulting in a successful double spending.

\section{Consensus or Broadcast}
Consensus is a property that allows multiple parties to reach agreement on transactions, either accepting or rejecting them. In the context of blockchain systems, consensus can be defined by the following set of properties:

\begin{definition}
Each honest validator observes some transaction from a set of conflicting transactions $\{t_0, t_1, \dots\}$.
\textbf{Consensus} satisfies the following properties:

\begin{description}
    \item \textbf{Totality}: If some honest validator accepts a transaction, every honest validator will eventually accept the same transaction.
    \item \textbf{Agreement}: No two honest validator accept conflicting transactions.
    \item \textbf{Validity}: If every honest validator observes the same transaction (there are no conflicting transactions), this transaction will be accepted by all honest validators.
    \item \textbf{Termination:} Some transaction from the set will eventually be accepted by honest validators.
\end{description}
\end{definition}





As implied by Agreement and Termination, a consensus protocol enables nodes to reach an agreement on conflicting transactions, where multiple valid transactions consuming the same input are involved.
In such cases, all nodes should unanimously accept one of the conflicting transactions.

As we have established, the Avalanche protocol features a relatively weak, sublinear resilience to liveness attacks involving conflicting transactions. To address this issue, Avalanche introduces the term of virtuous transactions, which can enjoy better guarantees.
In other words, even for a relatively small adversary, Avalanche does not satisfy the Termination property, and only guarantees termination if the Validity condition is also met: all honest validators observe just one valid transaction and no conflicting ones.

The termination property becomes crucial in scenarios involving smart contracts, where conflicting transactions may arise, such as two users attempting to purchase the same product.
To address this limitation, the Avalanche team introduced a different solution for the C-Chain and P-Chain, specifically designed to execute smart contracts required for such blockchain applications.

As described by \cite{consensusnumber}, consensus is not necessary for payment systems, and indeed there exist payment systems providing similar guarantees to Avalanche, while also unable to support general applications such as smart contracts: broadcast-based payment systems \cite{fastpay,astro,zef,abc,limitlessly}.
The provided guarantees of a Byzantine reliable broadcast can be defined as follows:

\begin{definition}
Each honest validator observes some transaction from a set of conflicting transactions $\{t_0, t_1, \dots\}$.
\textbf{Byzantine reliable broadcast} satisfies the properties of Consensus, without the Termination property.
\end{definition}

Thus, referring to Avalanche as a consensus protocol can be misleading, as it is more akin to broadcast-based payment systems. While the performance of Avalanche is given prominence, a different solution has been used as required by the C-Chain and P-Chain.


\section{Related Work}
Amores-Sesar et al.~\cite{amores2022spring} analyze the Avalanche protocol.
They explain the protocol with pseudocode and introduce a property of Avalanche that was omitted here:
No-op transactions which are stateless transactions are added into the DAG by the nodes to make sure we always make progress on the finalization of older transactions.
The paper introduces a liveness attack (different from ours) that could be possible if the naive way of voting for a transaction and its ancestors with just a binary yes/no vote was implemented.
However, this is not the case, as emphasized at the end of the paper with the pseudocode involving the virtuous frontier concept.

Ash Ketchum and Misty Williams~\cite{pseudo-profound_bullshit} raise concerns similar to ours in their recent write-up, that Avalanche is not a consensus protocol.

Most recently, a follow-up analysis by Amores-Sesar et al.~\cite{amores2024analysis}
formalized the need for at least $\Omega(\log n + \beta)$ rounds for consensus with the Snow family of protocols.
They then proposed a specific modification of Snowflake and Snowball implementing this change.

\section{Conclusion}
In this paper, we have examined the resilience properties of Avalanche and its underlying Snowball protocol. We have experimentally evaluated simple strategies for a potential adversary. To quantify the efficacy of these attacks, we have conducted simulations and evaluated the ratio of stake the adversary needs to control to launch successful attacks on liveness and safety.

Our analysis revealed that an adversary with a small fraction of the stake can indefinitely keep the network in a state where it cannot finalize a transaction.
With some probability depending on the stake and the size of the network, the adversary can also convince some node to finalize a transaction that is then rejected by other honest parties, which can result in a double spending attack.

Through our analysis, we have demonstrated that the Snowball protocol - the foundation of Avalanche - is vulnerable, when conflicting transactions are present.
The weak resilience when conflicting transactions are present is a critical limitation, as it makes the protocol unable to support general smart contracts.
This explains why Avalanche actually uses a different protocol, called Snowman, which uses a linear blockchain (instead of a DAG) in order to totally order those transactions, unlike what is done for payments~\cite{avalanche_platform}.

\subsubsection{Future Work}
The basis of our attacks relies on the presence of conflicting transactions.
Future work could analyze how Avalanche distinguishes unique transactions, and determine the feasibility for an adversary to arbitrarily create conflicting transaction from another transaction $T$ broadcast by an honest node, for example, by creating a copy with different parents in the DAG.

\label{lastpagebeforerefs}

\bibliographystyle{splncs04}
\bibliography{references}

\appendix

\end{document}